\documentstyle[12pt]{article}
\begin{document}

\title{Description of deformed nuclei in the sdg boson model}

\author{S.C. Li$^1$ and S. Kuyucak$^2$ \\
Department of Theoretical Physics,\\ Research School of Physical Sciences,\\
Australian National University,
Canberra, ACT 0200, Australia\\}
\date{}

\maketitle

\begin{abstract}
We present a study of deformed nuclei in the framework of the sdg interacting
boson model utilizing both numerical diagonalization and analytical $1/N$
expansion techniques. The focus is on description of high-spin states which
have recently become computationally accessible through the use of computer
algebra in the $1/N$ expansion formalism. A systematic study is made of 
high-spin states in rare-earth and actinide nuclei.
\end{abstract} 

\vskip 1cm
\noindent
$^1$ E-mail: scl105@rsphy1.anu.edu.au \\
$^2$ E-mail: sek105@phys.anu.edu.au
\vfill \eject

\baselineskip=21pt

\section{Introduction \label{int}}
Application of the interacting boson model (IBM) \cite{iac87} to deformed 
nuclei has been fraught with conceptual and technical difficulties
from the early days. Truncation of the model space to s and d bosons created
a lot of controversy and was challenged on both microscopic \cite{bm80} and
phenomenological \cite{bm82} grounds. When the dust finally settled, it was 
generally accepted that g bosons were both necessary and sufficient for 
microscopic foundation \cite{yos84} as well as phenomenological description
of deformed nuclei in the IBM \cite{yos88} (see also \cite{cas88,dev92,hey93}
for recent reviews). The sdg-IBM, however, does not share many of the
simplifying features of the sd-IBM which made it so popular. To start with, the
symmetry limits of the dynamical group SU(15) are not associated with actual
spectra so there is little help from group theory. Secondly, due to the large
basis spaces, exact diagonalization of sdg-IBM Hamiltonians is not possible for 
deformed nuclei. As a result, progress in sdg-IBM calculations has been
rather slow, and due to various approximations involved, a satisfactory
description of both low-lying band structures and high-spin states in deformed
nuclei is still missing. 

Numerical calculations in the sdg-IBM were initially performed by coupling 
a single g boson to an sd boson core \cite{isa82}. In another approach,
a Hamiltonian consisting of various SU(3) tensor operators was diagonalized 
in a truncated SU(3) basis \cite{yos88}. At present, full basis sdg-IBM 
calculations are possible only for vibrational and transitional nuclei with
boson numbers $N\le 10$ \cite{lac92}. The accuracy of the truncated space
calculations for deformed nuclei with $N\ge 12$ has not been explored so far.
One of the purposes of this paper is to examine the convergence properties of 
various matrix elements (m.e.) with the number of g bosons allowed, and to
establish some criteria for the accuracy of truncated sdg-IBM calculations. 
It should be clear that due to truncation, the validity of numerical
diagonalization results in the sdg-IBM is limited to low-lying states. 
Besides, such calculations are time consuming and therefore are not well suited
to explore the effect of the multitude of parameters in the sdg-IBM. 
The $1/N$ expansion method \cite{kuy88}, which is based on angular momentum
projected mean field theory, could alleviate both problems. The available
analytic formulas for various physical quantities allow systematic study of
model parameters, and fast and efficient analysis of data \cite{kuy91}. 
In addition, the $1/N$ expansion has recently been extended to higher orders
using computer algebra \cite{kuy95a}, thus enabling accurate descriptions of
high-spin states within the sdg-IBM. Currently the $1/N$ expansion formalism
covers ground- and single-phonon bands. Calculations for multi-phonon bands are 
rather complicated and have not been performed yet. Therefore, at this stage,
one has to rely on both analytical and numerical methods to obtain a complete
description of deformed nuclei 

In this paper, we first discuss the recent developments in the numerical
diagonalization and the 1/N expansion methods. We then present a systematic
study of deformed nuclei in the sdg-IBM employing a Hamiltonian consisting of
one-body energies and quadrupole and hexadecapole interactions. Application of
the results is focused on the high-spin states in rare-earth and actinide
nuclei as their description in the IBM has been a source of criticism
\cite{bm82} which has not been adequately addressed so far. In order to
constrain the model parameters properly, both the high-spin data and the
low-lying band structures are described simultaneously.

\section{Choice of Hamiltonian \label{hamiltonian}}
The consistent-$Q$ formalism (CQF) \cite{cas88} has been the standard choice
for description of deformed nuclei in the sd-IBM. The CQF Hamiltonian consists
of the dipole and quadrupole interactions
\begin{equation}
H = \kappa_1 L \cdot L - \kappa_2 Q \cdot Q,
\label{hamsd}
\end{equation}
where the quadrupole operator is given by
\begin{equation}
Q=[s^\dagger {\tilde d} + d^\dagger {\tilde s}]^{(2)} +
\chi [d^\dagger {\tilde d}]^{(2)}.
\label{qsd}
\end{equation}
Here brackets denote tensor coupling of the boson operators and
$\tilde b_{l\mu}=(-1)^{\mu}b_{l-\mu}$.
For consistency, the same quadrupole operator is used in the E2 transition
operator, $T$(E2)$=e_2 Q$, as in the Hamiltonian. The CQF has been especially
successful in reproducing the low-lying band structures and the E2 transitions 
among them \cite{cas88}. A more complete description of deformed nuclei,
including high-spin states and hexadecapole bands, however, requires extension
of the model space to sdg bosons.

A minimal extension of the CQF Hamiltonian to the sdg-IBM can be achieved by
including the g-boson energy term, $\varepsilon_g \hat n_g$ in the Hamiltonian
(\ref{hamsd}), and modifying the quadrupole operator (\ref{qsd}) to
\begin{equation} 
Q=[s^\dagger {\tilde d} + d^\dagger {\tilde s}]^{(2)} +
q_{22} [d^\dagger {\tilde d}]^{(2)} + q_{24} [d^\dagger {\tilde g} +
g^\dagger {\tilde d}]^{(2)} + q_{44} [g^\dagger {\tilde g}]^{(2)}.
\label{qsdg}
\end{equation}
We shall refer to this minimal extension as the CQF below.
A study of high-spin states in the sdg-IBM using the CQF Hamiltonian has
indicated that the energy surface remains too rigid, and inclusion
of the d-boson energy term, $\varepsilon_d \hat n_d$, is essential in order to 
reproduce the spin dependence of moment of inertia (MOI) and E2 transitions in
the ground band \cite{kuy95b}. The success of this pairing plus quadrupole type
of Hamiltonian, however, does not extend to the side bands which are more
sensitive to interference from the hexadecapole interaction.
Thus, for a comprehensive description of deformed nuclei one needs to employ 
the Hamiltonian
\begin{equation}
H= \varepsilon_d \hat n_d + \epsilon_g \hat n_g - \kappa_2 Q \cdot Q
- \kappa_4 T_4 \cdot T_4,
\label{hamsdg}
\end{equation}
where the hexadecapole operator is given by
\begin{equation}
T_4=[s^\dagger {\tilde g} + g^\dagger {\tilde s}]^{(4)} +
h_{22} [d^\dagger {\tilde d}]^{(4)} + h_{24} [d^\dagger {\tilde g} +
g^\dagger {\tilde d}]^{(4)} + h_{44} [g^\dagger {\tilde g}]^{(4)}.
\label{hexop}
\end{equation}
Note that we have deliberately left out the dipole interaction in 
eq.~(\ref{hamsdg}) as it is at the root of the rigid MOI problem in the IBM. 
This Hamiltonian contains 10 parameters, namely, the one-body energies
$\varepsilon_d$ and $\varepsilon_g$, the multipole interaction strengths 
$\kappa_2$ and $\kappa_4$, and the quadrupole and hexadecapole parameters
$q_{jl}$ and $h_{jl}$. In a systematic study covering many nuclei, it is
desirable to have a smaller set of free parameters. To achieve this goal, we
adapt a similar strategy as in a previous study of deformed nuclei
\cite{kuy91}. The quadrupole parameters $\{q_{22},q_{24},q_{44}\}$ are scaled
from their SU(3) values with a single factor $q$ as suggested by microscopics
\cite{oai78}, while the hexadecapole parameters $h_{jl}$ are determined from
those of $q_{jl}$ through the commutation condition, $[\bar{h},\bar{q}]=0$,
which ensures that the quadrupole and the hexadecapole mean fields are coherent
\begin{equation}
\bar{h}_{22}=\bar{q}_{24},\,\; \bar{h}_{24}=\bar{q}_{44}, \, \;
\bar{h}_{44}=\bar{q}_{24}+(\bar{q}_{44}^2-\bar{q}_{22}\bar{q}_{44}-1)/
\bar{q}_{24}\;.
\label{hexpar}
\end{equation}
Here $\bar{q}_{jl}=\langle j0l0|20\rangle q_{jl}$ and 
$\bar{h}_{jl}=\langle j0l0|40\rangle h_{jl}$.
This reduction of parameters from 10 to 5 is obtained at the expense of
detailed description of quadrupole and hexadecapole operators. Since
information, especially on the latter, is rather patchy, this will not 
cause any problems except in a few isolated cases.
In the calculation of E2 and E4 transitions, we shall use the 
consistent operators, $T$(E2)$=e_2 Q$, $T$(E4)$=e_4 T_4$, so that, apart from 
the effective charges $e_2$ and $e_4$, no new parameters are introduced.

In concluding this section, we present an alternative parametrization of
the Hamiltonian (\ref{hamsdg}) which is more convenient in implementation of
the $1/N$ expansion formulas. This involves factoring out the energy scale and
the leading order $N$ dependence from the energy expressions. 
Since the quadrupole interaction is dominant, a suitable choice for such a set 
of dimensionless parameters is given by
\begin{equation}
\eta_l = \varepsilon_l/N\kappa_2, \quad \zeta_k=\kappa_k/\kappa_2,
\label{par}
\end{equation}
where $l=0,2,4, \dots$ correspond to the subscripts $s,d,g, \dots$.

\section{Numerical diagonalization \label{numerical}}
A computer code which can diagonalize arbitrary sdg-IBM Hamiltonians in
full space has been available for some time \cite{mor86}, but due to excessive
memory requirements it had only limited applications to transitional nuclei
\cite{lac92}. Recently this code has been modified to run on supercomputers 
improving its applicability \cite{li93}. Nevertheless, exact diagonalization
for deformed nuclei with $N > 10$ remains elusive due to the large basis
space, and truncation of the model space is still necessary.
Here we use the modified code to study the accuracy of the truncated space
calculations in the sdg-IBM. Numerical results will also be used in
demonstrating the accuracy of the $1/N$ expansion formulas in sect. \ref{1N},
and in application to actual spectra in sect. \ref{application}.

Since g bosons are relatively weakly coupled, a natural parameter in 
truncating the basis space is the maximum number of g bosons allowed, 
$n_{g\max}$. In figs. 1-2, we examine the convergence properties of some key
observables as a function of $n_{g\max}$. For this purpose we use a typical
deformed Hamiltonian with parameters (as defined in eq.~(\ref{par})),
$\kappa_2=-20$ keV, $\kappa_4=0$, $q=0.5$, $\eta_d=1.5$, $\eta_g=4.5$, and 
$N=10$. Fig.~1 shows the effect of the truncation on low-lying band structure:
a) band excitation energies, b) E2 transitions, and c) E4 transitions. 
The $n_{g\max}=1$ calculations are off by about 10-20\% (mostly overestimated
but underestimated in a few cases), and hence they are not very reliable. 
As expected, the hexadecapole bands take longer to converge 
compared to the $\beta$ and $\gamma$ bands, the worst case being the
$\beta'$ band. Nevertheless, convergence to accuracy of a few percent is
obtained in almost all cases for $n_{g\max}=3$. 
In fig.~2, we present a similar study for the high-spin states in the ground 
band: a) excitation energies, and b) E2 transitions. At spins $L\sim 2N$,
the $n_{g\max}=1$ calculations are off by about 20-30\% which will get even
worse with increasing spin. The $n_{g\max}=3$ results, on the other hand,
provide a reasonably accurate picture up to spins $L\sim 2N$. Beyond that,
g bosons start dominating the wave functions, and any truncation is likely to
lead to substantial errors. 

The above results suggest that diagonalization of the sdg-IBM Hamiltonians
in a model space truncated to $n_{g\max}=N/3$ bosons will give a 
reliable description of states with spins $L<2N$. This extends the
applicability of the SDGBOSON code to $N=14$ which covers roughly half of the 
deformed nuclei. It should nevertheless be emphasized that these computations
are expensive, time consuming, and certainly not the best way to deal with
the sdg-IBM problems. In the next section, we introduce the $1/N$ expansion 
which circumvents the shortcomings of numerical diagonalization.

\section{$1/N$ expansion formalism for high-spin states \label{1N}}
The $1/N$ expansion formalism \cite{kuy88} was developed as a response to
difficulties in performing calculations in the sdg-IBM due to the 
inadequacy of group theoretical techniques and the large basis space problem 
in numerical diagonalization. It is based on angular momentum projected 
mean field theory and leads to analytic expressions for various physical
quantities of interest. Initially, the $1/N$ calculations were carried out to
order $1/N^2$ which is quite sufficient for low-lying states.
An accurate description of high-spin states, on the other hand, requires
inclusion of terms up to order $1/N^6$ which is not suitable for hand 
calculation. This difficulty has finally been overcome through the use of
computer algebra. A brief description of the method for the
ground band was given previously \cite{kuy95a}. Here we provide details 
of the extended calculations including the results for the excited bands and 
the electromagnetic transitions.

\subsection{Ground band}
We consider a general formulation of the IBM as this allows an elegant
derivation of the $1/N$ formulas by fully exploiting the angular momentum 
algebra. Thus we introduce the boson creation and annihilation operators 
$b^\dagger_{l\mu}, b_{l\mu}$ with $l=0,2,4, \ldots$.\ where $b_0=s$, 
$b_2=d$, $b_4=g$, etc. 
In order to keep the variational problem to a manageable size, it is 
necessary to assume that the boson system is axially symmetric, and hence $K$ 
is a good quantum number. From comparison with the exact diagonalization
results, this assumption will be seen to hold to a very good degree.
The ground band then can be written as a condensate of intrinsic bosons as
\begin{equation}
|\phi_g\rangle =(N!)^{-1/2}(b^\dagger)^N|0\rangle,\quad
b^\dagger=\sum_{l} x_{l} b^\dagger_{l0},
\label{ground}
\end{equation}
where $x_{l}$ are the normalized boson mean fields, i.e.
${\bf x}.{\bf x}=1$, ${\bf x} = (x_0, x_2, x_4, \dots)$. 
In the classical limit of the IBM, the mean fields are associated with the
deformation parameters of the system \cite{gin80}. For a given Hamiltonian $H$,
they are determined from $\langle H \rangle_L$ by variation after projection
(VAP). 

The Hamiltonians in eqs. (\ref{hamsd},\ref{hamsdg}) can be written
in the generalized form as
\begin{equation}
H = \sum_l \varepsilon_l \hat n_l - \sum_{k=0}^{2l_{\max}}
\kappa_k T^{(k)}\cdot T^{(k)}, \quad
\hat n_l = \sum_\mu b^\dagger_{l\mu} b_{l\mu}, \quad
T^{(k)} = \sum_{jl} t_{kjl}[b_j^\dagger \tilde b_l]^{(k)},
\label{ham}
\end{equation}
where the parameters have the obvious correspondence,
$\varepsilon_2=\varepsilon_d$, $\varepsilon_4=\varepsilon_g$, $t_{2jl}=q_{jl}$,
$t_{4jl}=h_{jl}$. 
This general form has the advantage that, to evaluate the expectation
value of $H$, one needs to perform the calculation for a generic number 
operator $\hat n_l$ and a multipole interaction $T^{(k)}\cdot T^{(k)}$.
The expectation value of a scalar operator $\hat O$ in the ground band 
(\ref{ground}), with angular momentum projection, is given by
\begin{equation}
\langle \hat O \rangle_L = {2L+1 \over 2N! {\cal N}(\phi_g,L)}
\int d \beta \sin\beta \, d^L_{00}(\beta) 
\langle 0|b^N \hat O e^{-i \beta L_y} (b^\dagger)^N |0 \rangle.
\label{nl1}
\end{equation}
Here, the normalization, ${\cal N}(\phi_g,L)$, follows from eq.~(\ref{nl1})
upon substituting the identity operator for $\hat O$. 
Algebraic manipulations in eq.~(\ref{nl1}) are most easily carried out using
boson calculus and angular momentum algebra techniques 
(see ref. \cite{kuy88} for a pedagogical treatment).
For the number operator, one obtains
\begin{equation}
\langle\hat n_l\rangle_L = {N x_l^2 \over F(N,L)}
\sum_I \langle L0l00|I0\rangle^2 F(N-1,I),
\label{nl2}
\end{equation}
where $F(N,L)$ denotes the reduced normalization integral
\begin{equation}
F(N,L) = {\cal N}(\phi_g,L)/(2L+1).
\label{norm1}
\end{equation}
Eq.~(\ref{nl2}) is exact and highlights the essential role played by the 
normalization integral. In the original papers \cite{kuy88}, a Gaussian 
approximation was used in the evaluation of $F(N,L)$ which limited the accuracy 
of m.e. to order $1/N^2$. This difficulty has been overcome recently 
using the computer algebra software Mathematica \cite{wol91}. By exploiting 
the symmetries of the boson system, the normalization integral is cast into a
system of linear equations which is solved with the help of Mathematica
\cite{kuy95c}. The result is a double expansion in $1/N$ and $\bar L=L(L+1)$
given by 
\begin{equation}
F(N,L)={2\over aN} \sum_{n=0} {(-1)^n\over n!(aN)^n}
\sum_{m=0}^n \alpha_{nm} \bar L^m.
\label{norm2}
\end{equation}
The coefficients $\alpha_{nm}$ in eq.~(\ref{norm2}) are given in terms of
polynomials of the moments of $x_l^2$ 
\begin{equation}
a_n = \sum_l \bar l^{n+1} x_l^2,
\label{an}
\end{equation}
and $a$ is defined as $a \equiv a_0$.
A list of $\alpha_{nm}$ up to sixth order is given in ref. \cite{kuy95c}.
The knowledge of $F(N,L)$, in principle, allows evaluation of m.e.
to arbitrary orders in $1/N$. As will be seen in the applications, a correct
description of MOI at high-spins requires inclusion of 
terms of order $\bar L^3/N^6$. Evaluation of eq.~(\ref{nl2}) to such
high orders is too difficult to perform by hand but becomes manageable
using computer algebra. 

Before presenting the final results, it will be useful to comment on the
general form of the m.e. of a $k$-body operator $\hat O$, and illustrate
the concept of layers in the 1/$N$ expansion
\begin{eqnarray}
\langle\hat O\rangle_L = N^k \sum_{n,m} {O_{nm}\over (aN)^m}
\Bigl({\bar L \over a^2N^2}\Bigr)^n \hskip 1.9cm & &\nonumber \\[.2cm]
 = N^k \Bigl\{ O_{00}+{O_{01}\over aN}+{O_{02}\over (aN)^2}
+{O_{03}\over (aN)^3}&+& \cdots \nonumber \\
+{\bar L\over a^2N^2}\ \Bigl(\ O_{10}\ +\ \ {O_{11}\over aN}
+{O_{12}\over (aN)^2}
&+& \cdots \Bigr) \nonumber \\
+\Bigl({\bar L\over a^2N^2}\Bigr)^2 \
\Bigl(\ O_{20}\ +\ \ {O_{21}\over aN}&+& \cdots \Bigr) \nonumber \\
+\Bigl({\bar L\over a^2N^2}\Bigr)^3\ \Bigl(\ O_{30} &+& \cdots \Bigr)
+ \cdots \Bigr\}.
\label{me1}
\end{eqnarray}
The expansion coefficients $O_{nm}$ in eq.~(\ref{me1}) involve various
quadratic forms of the mean fields $x_{l}$ corresponding to the single-boson
m.e. of $\hat O$ and its moments. The explicit form is given to facilitate the
illustration of layers. Notice that the $i$ coefficients
$O_{nm}$ in the $i$'th column have $n+m=i-1$ constant, and are referred as the
layer ``$i-1$". The leading term in eq.~(\ref{me1}) thus forms the zeroth
layer. (This name is appropriate since calculations in the intrinsic frame give
the same result independent of projection.) In simple terms, the layer 
in an expansion is given by the maximum power of $\bar L$. There is a close
connection between the layers in the m.e. (\ref{me1}) and the normalization
coefficients $\alpha_{nm}$ in eq.~(\ref{norm2}), namely, in order to calculate
the m.e. up to the $i$'th layer, one needs to know the coefficients
$\{\alpha_{nn}, \alpha_{nn-1}, \dots, \alpha_{nn-i+1}, n=1,2i\}$. This is very
useful in higher order calculations as it restricts the number of terms in the
expansion, cutting down the amount of algebra. To make this point clear, we
note that eq.~(\ref{me1}) shows all the terms in the third layer whereas a
complete calculation to order $1/N^6$ would require 6 more terms belonging to
the fourth, fifth and sixth layers. As can be seen from eq.~(\ref{nl3}) below,
the complexity of the coefficients $O_{nm}$ increases ``exponentially" with
layers, and each of the extra terms would lead to expressions pages long. From
a practical point of view, such accuracy is never required. The only $1/N^6$
term of any consequence is $\bar L^3/N^6$ which is included in the third layer.
The rest are completely negligible. Hence use of layers is a more sensible
approach than a complete calculation to a given order in $1/N$. 

With these considerations, we present the result of the Mathematica evaluation
of the one-body m.e. (\ref{nl2}) to the third layer 
\begin{eqnarray}
&&\hskip -1cm \langle\hat n_l\rangle_L= Nx_l^2 \Bigl\{
1 + {1 \over aN}\Bigl(a-\bar l\Bigr)
+ {1 \over (aN)^2}\Bigl(-a + a_1/2 + (1 - a_1/a)\bar l + \bar l^2/2\Bigr)
\nonumber\\
&&\hskip 1cm + {1 \over (aN)^3}\Bigl(
a + 2a^2 - 7a_1/3 - aa_1 + 5a_1^2/4a - a_2/3 \nonumber\\
&&\hskip 2.2cm +  (-1 - 2a + 2a_1+ 7a_1/2a- 5a_1^2/2a^2 + a_2/2a)\bar l
\nonumber\\
&&\hskip 2.2cm + (-7/6 - a + 5a_1/4a)\bar l^2 - \bar l^3/6 \Bigr) \nonumber\\
&&+ {\bar L \over (aN)^2}\Bigl[ (-a+\bar l)
+ {1 \over aN}\Bigl(2a+2a^2-2a_1+(-2-2a+3a_1/a)\bar l-\bar l^2 \Bigr)
\nonumber\\
&&\hskip 1cm + {1 \over (aN)^2}\Bigl(-3a-12a^2-4a^3+21a_1/2+11aa_1
-15a_1^2/2a+3a_2/2  \nonumber \\
&&\hskip 2.2cm  + (3+12a+4a^2-33a_1/2-14a_1/a+25a_1^2/2a^2-2a_2/a)\bar l
\nonumber \\
&&\hskip 2.2cm  + (7/2+11a/2-5a_1/a)\bar l^2 + \bar l^3/2 \Bigr) \Bigr]
\nonumber \\
&&+ {\bar L^2\over 2(aN)^4}\Bigl[
(-a-2a^2+3a_1/2+(1+2a-2a_1/a)\bar l+\bar l^2/2) \nonumber\\
&&\hskip 1cm + {1 \over aN}\Bigl( 4a+21a^2+14a^3-16a_1-51aa_1/2+13a_1^2/a-2a_2
\nonumber\\
&&\hskip 2.2cm +   (-4-21a-14a^2+34a_1+20a_1/a-39a_1^2/2a^2+5a_2/2a)\bar l
\nonumber\\
&&\hskip 2.2cm + (-4-17a/2+13a_1/2a)\bar l^2-\bar l^3/2 \Bigr) \Bigr]
\nonumber\\
&&+ {\bar L^3\over 3(aN)^6}\Bigl[
-a-6a^2-6a^3+25a_1/6+9aa_1-15a_1^2/4a+5a_2/12 \nonumber\\
&&\hskip 2.2cm +  (1+6a+6a^2-45a_1/4-5a_1/a+21a_1^2/4a^2-a_2/2a)\bar l
\nonumber\\
&&\hskip 2.2cm + (5/6+9a/4-3a_1/2a)\bar l^2+\bar l^3/12 \Bigr] \Bigr\},
\label{nl3}
\end{eqnarray}
where $a_n$ is defined in eq.~(\ref{an}). 
Eq.~(\ref{nl3}) can be checked against two results: i) it satisfies the
number conservation, i.e. $\sum_l \langle \hat n_l \rangle_L =N$, and
ii) it reproduces the analytic formulas available in the SU(3) limit
\cite{iac87}. 

A similar calculation for the multipole interaction yields the intermediate 
result
\begin{eqnarray}
&&\hskip -1cm \langle T^{(k)}\cdot T^{(k)}\rangle_L = {N(2k+1)\over F(N,L)}
\biggl\{ \sum_{jl} {(t_{kjl} x_l)^2 \over 2l+1}
\sum_I \langle L0l0|I0\rangle^2 F(N-1,I) \nonumber\\
&&\hskip 4.cm + (N-1) \sum_{jlj'l'J} t_{kjl} t_{kj'l'} x_j x_l x_{j'} x_{l'}
\langle j0j'0|J0\rangle \langle l0l'0|J0\rangle \nonumber\\
&&\hskip 4.cm \times
\left\{ \begin{array}{ccc} j & j' & J\\l' & l & k\end{array}\right\}
\sum_I \langle L0J0|I0\rangle^2 F(N-2,I) \biggr\}.
\label{tk1}
\end{eqnarray}
Again this is exact and can be evaluated to any order using Mathematica. 
The third layer result is given by
\begin{eqnarray} 
&&\hskip -1cm \langle T^{(k)}\cdot T^{(k)}\rangle_L=
N^2 \Bigl\{ U_k + {1 \over aN}\Bigl(aU_k - U_{k1} + aC_k\Bigr) \nonumber\\
&&\hskip 2.3cm + {1 \over (aN)^2}\Bigl((-2a+a_1)U_k + (1-a-a_1/a)U_{k1} +
    U_{k2}/2 + a^2C_k-aC_{k1}\Bigr) \nonumber\\
&&\hskip 2.3cm + {1 \over (aN)^3}\Bigl((2a + 2a^2 - 14a_1/3 - aa_1+5a_1^2/2a-
2a_2/3)U_k\nonumber\\
&&\hskip 3.7cm + (-1 + a - a_1/2 + 7a_1/2a - 5a_1^2/2a^2+a_2/2a)U_{k1}
\nonumber\\
&&\hskip 3.7cm + (-7/6 + 5a_1/4a)U_{k2} - U_{k3}/6 \nonumber\\
&&\hskip 3.7cm + (-a^2 + aa_1/2)C_k + (a - a_1)C_{k1} + aC_{k2}/2\Bigr)
\nonumber\\
&&\quad + {\bar L \over (aN)^2}\Bigl[ -2aU_k + U_{k1} \nonumber\\
&&\hskip 1.7cm +  {1 \over aN}\Bigl((4a + 2a^2 - 4a_1)U_k
+ (-2 + a + 3a_1/a)U_{k1} - U_{k2} - a^2C_k + aC_{k1}\Bigr)  \nonumber\\
&&\hskip 1.7cm+ {1 \over (aN)^2}\Bigl((-6a - 16a^2 - 4a^3 + 21a_1 + 15aa_1 -
      15a_1^2/a + 3a_2)U_k \nonumber\\
&&\hskip 3.2cm + (3+2a-2a^2-4a_1-14a_1/a + 25a_1^2/2a^2 - 2a_2/a)U_{k1}
\nonumber\\
&&\hskip 3.2cm + (7/2+2a-5a_1/a)U_{k2}+U_{k3}/2 \nonumber\\
&&\hskip 3.2cm + (2a^2 + 2a^3-2aa_1)C_k+(-2a-2a^2+3a_1)C_{k1} -aC_{k2}\Bigr)
\Bigr] \nonumber\\
&&\quad  + {\bar L^2\over 2(aN)^4}\Bigl[ (-2a - 2a^2 + 3a_1)U_k +
(1 - 2a_1/a)U_{k1} + U_{k2}/2 \nonumber\\
&&\hskip 2.cm+ {1 \over aN}\Bigl((8a + 30a^2 + 14a^3 - 32a_1 - 37aa_1 +
               26a_1^2/a - 4a_2)U_k \nonumber\\
&&\hskip 3.2cm + (-4 - 8a + 2a^2 + 29a_1/2 + 20a_1/a - 39a_1^2/2a^2 +
                  5a_2/2a)U_{k1} \nonumber\\
&&\hskip 3.2cm + (-4-9a/2 + 13a_1/2a)U_{k2} - U_{k3}/2 \nonumber\\
&&\hskip 3.2cm + (-a^2 - 2a^3 + 3aa_1/2)C_k + (a + 2a^2 - 2a_1)C_{k1}
+ aC_{k2}/2\Bigr) \Bigr] \nonumber\\
&&\quad + {\bar L^3\over 3(aN)^6}\Bigl[ (-8a - 36a^2 - 24a^3 + 100a_1/3
+ 54aa_1 - 30a_1^2/a + 10a_2/3)U_k \nonumber\\
&&\hskip 3.2cm + (4 + 12a - 24a_1 - 20a_1/a + 21a_1^2/a^2 - 2a_2/a)U_{k1}
\nonumber\\
&&\hskip 3.2cm + (10/3 + 6a - 6a_1/a)U_{k2} + U_{k3}/3\Bigr] \Bigr\}.
\label{tk2}
\end{eqnarray}
Here the quadratic forms $C_{kn}$ arise from normal ordering and simulate
an effective one-body term
\begin{equation}
\quad C_{kn} = (2k+1) \sum_{jl} \bar{l}^n (t_{kjl} x_l)^2/(2l+1),
\end{equation}
while $U_{kn}$ represent the genuine two-boson interaction
\begin{equation}
U_{kn}=\sum_{jlj'l'I} \bar I^n \langle j0j'0|I0\rangle \langle l0l'0|I0\rangle
\left\{ \begin{array}{ccc} j & j' & I\\l' & l & k\end{array}\right\}
t_{kjl} t_{kj'l'} x_j x_l x_{j'} x_{l'}.
\end{equation}
For a given multipole, these sums can be evaluated in closed form using
Mathematica. For the quadrupole and hexadecapole interactions, the first four
terms needed in eq.~(\ref{tk2}) are given by 
\begin{eqnarray}
&&U_2  = A^2,\nonumber\\
&&U_{21} = (2A_1-3A)A,\nonumber\\
&&U_{22} = (2A_2-24A_1+18A)A + (A_{11}-A_2+7A_1)A_1 + (A_{11}-A_2)^2/12,
\nonumber\\
&&U_{23} = (2A_3-36A_2-18A_{11}+240A_1-144A)A \nonumber\\
&&\hskip 1cm +(3A_{21}-3A_3+56A_2+16A_{11}-194A_1)A_1/2\nonumber\\
&&\hskip 1cm +(11A_{11}^2+14A_{11}A_2- 25A_2^2)/12
+(A_3-A_{21})(A_2-A_{11})/4, \nonumber\\
&&U_4  = B^2,  \nonumber\\
&&U_{41} = (2B_1 - 10B)B/2, \nonumber\\
&&U_{42} = 4B_1^2 + (2B_{11} - 40B_1 + 20B)B 
       + (B_2  - B_{11}- 20B_1 +180B )^2/180 \nonumber\\
&&U_{43} = (2B_3-120B_2-60B_{11}+2480B_1-4400B)B \nonumber\\
&&\hskip 1cm + (3B_{21}-3B_3+224B_2+58B_{11}-2756B_1/3)B_1/9 \nonumber\\
&&\hskip 1cm + (8B_{11}^2+14B_{11}B_2-11B_2^2)/45 + (B_3-B_{21})(B_2-B_{11})/60
\label{ukn}
\end{eqnarray}
Here the quadratic forms $A_{mn}$ and $B_{mn}$ in (\ref{ukn}) are defined as
\begin{eqnarray}
A_{mn}=\sum_{jl} \bar{j}^m \bar{l}^n \langle j0l0|20\rangle t_{2jl} x_j x_l,
\qquad
B_{mn}=\sum_{jl} \bar{j}^m \bar{l}^n \langle j0l0|40\rangle t_{4jl} x_j x_l,
\label{amn}
\end{eqnarray}
and correspond to various moments of the single-boson m.e. of the quadrupole
and hexadecapole operators. (note that the zero subscripts are suppressed for
convenience). The quadrupole m.e. given by (\ref{tk2}-\ref{ukn}) reproduces the
well known Casimir eigenvalues in the SU(3) limit, hence also passes the SU(3)
test. 

The analytic expressions presented above are already rather long.
If for any reason, the next layer results should be required, the expressions
would grow to pages long, and the analytical $1/N$ calculations might not be
very practical.
In such cases, numerical evaluation of the m.e. (\ref{nl2},\ref{tk1}),
as described in Appendix A, may be preferable. Although this would increase 
the computation time appreciably, it has the advantage that the calculations 
are done exactly to all orders in $1/N$.

\subsection{Variation after projection \label{vap}}
The energy expression derived in the last subsection is rather lengthy,
and in discussing the variational problem, it will be more convenient to
express it in a compact form. Thus, using the parametrization in 
eq.~(\ref{par}), we rewrite the ground band energy as 
\begin{equation} 
E_{gL} =  N^2 \kappa_2 \sum_{n,m} {E_{nm}\over N^m} 
\Bigl({\bar L \over N^2}\Bigr)^n,
\label{eg}
\end{equation}
where the coefficients $E_{nm}$ can be read off from eqs. 
(\ref{nl3}-\ref{amn}). For example the leading order is given by
\begin{equation}
E_{00} = \sum_l \eta_l {x_l^2 \over {\bf x}.{\bf x}} 
- \left({A \over {\bf x}.{\bf x}}\right)^2 
- \zeta_4 \left({B \over {\bf x}.{\bf x}}\right)^2,
\end{equation}
with $A$ and $B$ defined in eq.~(\ref{amn}), and we have restored the
normalization factors ${\bf x}.{\bf x}$ as a precursor to variation. 
The minimum of the ground energy is obtained from 
\begin{equation}
\partial E_{gL} / \partial x_l = 0, \quad l = 0, 2, 4, \dots,
\label{der}
\end{equation}
which can be solved algebraically using the ansatz
\begin{equation}
{\bf x} = \sum_{n,m} {{\bf x}_{nm}\over N^m} \Bigl({\bar L \over N^2}\Bigr)^n.
\label{x}
\end{equation}
The use of layers again simplifies solution of the variational equations.
For the leading order (zeroth layer), one has the usual Hartree-Bose equations
\begin{equation}
\left.{\partial E_{00} \over \partial x_l}\right\vert_{{\bf x}_{00}} = 0,
\label{lay0}
\end{equation}
which are a system of coupled non-linear equations, and they are solved
numerically by iteration \cite{kuy88b}. Having determined ${\bf x}_{00}$, the
first layer mean fields ${\bf x}_{01}$ and ${\bf x}_{10}$ are then obtained by
solving the respective sets of equations 
\begin{eqnarray}
\left. {\partial E_{00} \over \partial x_l}\right\vert_{{\bf x}_{00}+ 
{\bf x}_{01}/N} 
&=& -{1 \over N} \left. {\partial E_{01} \over \partial x_l} 
\right\vert_{{\bf x}_{00}},
\nonumber\\
\left. {\partial E_{00} \over \partial x_l}\right\vert_{{\bf x}_{00}+
{\bf x}_{10} \bar L/N^2} 
&=& -{\bar L \over N^2} \left. {\partial E_{10} \over \partial x_l}
\right\vert_{{\bf x}_{00}}.
\label{lay1}
\end{eqnarray}
Upon substituting the mean fields in derivatives in (\ref{lay1}), the 
leading order vanishes by virtue of the Hartree-Bose eqs. (\ref{lay0}), and
the next order leads to sets of linear equations for ${\bf x}_{01}$ and 
${\bf x}_{10}$, that can be easily solved using Mathematica.
The Hartree-Bose condition also ensures that when the first layer mean
fields are substituted in the energy expression, the correction to the first
layer exactly vanishes. So they only contribute to the second and higher
layers \cite{kuy88b}. This holds in general for all layers. Thus for the third
layer expansion considered here, one needs at most the second layer mean fields 
${\bf x}_{02}$,  ${\bf x}_{11}$ and ${\bf x}_{20}$ which are obtained from
\begin{eqnarray}
\left. {\partial E_{00} \over \partial x_l}\right\vert_{{\bf x}_{00}+
{\bf x}_{01}/N+{\bf x}_{02}/N^2}
&=& -{1 \over N} \left. {\partial E_{01} \over \partial x_l}
\right\vert_{{\bf x}_{00}+{\bf x}_{01}/N} 
- {1 \over N} \left. {\partial E_{02} \over \partial x_l}
\right\vert_{{\bf x}_{00}},
\nonumber\\
\left. {\partial E_{00} \over \partial x_l}\right\vert_{{\bf x}_{00}+
{\bf x}_{10}\bar L/N^2+{\bf x}_{20} \bar L^2/N^4}
&=& -{\bar L \over N^2} \left. {\partial E_{10} \over \partial x_l}
\right\vert_{{\bf x}_{00}+{\bf x}_{10}\bar L/N^2}
- {\bar L^2 \over N^4} \left. {\partial E_{20} \over \partial x_l}
\right\vert_{{\bf x}_{00}},
\nonumber\\
\left. {\partial E_{00} \over \partial x_l}\right\vert_{{\bf x}_{00}+
{\bf x}_{01}/N+{\bf x}_{10} \bar L/N^2+{\bf x}_{11} \bar L/N^3}
&=& -{1 \over N} \left. {\partial E_{01} \over \partial x_l}
\right\vert_{{\bf x}_{00}+{\bf x}_{10} \bar L/N^2} 
-{\bar L \over N^2} \left. {\partial E_{10} \over \partial x_l}
\right\vert_{{\bf x}_{00}+{\bf x}_{01}/N}
\nonumber\\
&&-{\bar L \over N^3} \left. {\partial E_{11} \over \partial x_l}
\right\vert_{{\bf x}_{00}}, 
\label{lay2}
\end{eqnarray}
Again these sets of linear equations can be solved using Mathematica. 
We refrain from presenting these rather bulky results for the first and second 
layer mean fields here because, in the absence of analytical solutions for the
zeroth layer, they are not very illuminating. Upon substituting eq.~(\ref{x})
in (\ref{eg}), one obtains the variational corrections introduced by
the higher order mean fields in the ground band energies. These
lengthy analytic expressions contribute only to the second and higher layers 
and will not be shown here. All these results, together with
other $1/N$ expansion formulas, are nevertheless available in the form of a
Fortran code \cite{kl95}. Finally, if one is interested only in practical
applications of the results to high-spin states, one can determine the
minimum directly from the energy expression (\ref{eg}) using the numerical
simplex method, and thereby avoid the complexities introduced by the higher
order terms in the solution of the variational problem.

\subsection{Single-phonon bands}
Most of the high-spin data, as well as their theoretical analysis, are 
concentrated on the yrast bands (ground or two-quasiparticle).
While relying solely on the yrast data may be tolerated for microscopic models,
it could easily lead to misleading results in phenomenological models.
For this reason, inclusion of single-phonon bands in the analysis of  high-spin
data is highly desirable in phenomenological approaches. 
As will be seen in the applications, there are substantial high-spin data for
the $\gamma$ bands, which can be singled out among the single-phonon bands in
this respect. Therefore, we consider here the $\gamma$ band as an example of 
a single-phonon band calculation. Energy expressions for the other 
bands can be derived in a similar fashion. 

The single-phonon bands are obtained from the ground band by acting with the 
other intrinsic boson operators $b^\dagger_m=\sum_l x_{lm} b^\dagger_{lm}$, 
and then orthogonalizing the resulting bands. 
For example, the $\gamma$ band intrinsic state is given by 
\begin{equation}
|\phi_\gamma\rangle  = b^\dagger_2 |\phi_g, N-1\rangle + {1 \over
\sqrt{N-1}} \xi_\gamma (b^\dagger_{1})^2 |\phi_g,N-2 \rangle,
\label{gam}
\end{equation}
In this trial state, the mean fields for the ground band are already
established in the last subsection, and those for $b_1$ are determined from the
spurious $K=1$ band as \cite{kuy88}, 
\begin{equation}
x_{l1}=[\bar l/a]^{1/2} x_l.
\end{equation}
Thus, only the $\gamma$ band mean fields, $x_{l2}$, are to be determined by 
VAP. The coefficient $\xi_\gamma$ in eq.~(\ref{gam}) follows from the
orthogonality condition $\langle L_\gamma | L_g \rangle = 0$ as 
\begin{eqnarray}
&& \sum_{lI} \langle L2l-2|I0 \rangle \langle L0l0|I0 \rangle
\biggl[ x_{l} x_{l2} F(N-1,I) \nonumber\\
&&\hskip 1cm + \xi_\gamma \sum_{jj'} x_j x_{j1} x_{j'} x_{j'1} 
\langle j0j'0|l0 \rangle \langle j1j'1|l2 \rangle F(N-2,I) \biggr] = 0
\label{xi}
\end{eqnarray}
where $F$ denotes the ground band normalization (\ref{norm2}) for $N-1$
and $N-2$ bosons.

The expectation value of a scalar operator $\hat O$ in the $\gamma$ band
(\ref{gam}), with angular momentum projection, is given by 
\begin{eqnarray} 
&&\hskip -1cm \langle \hat O \rangle_{\gamma,L} = 
{2L+1 \over 2(N-1){\cal N}(\phi_{\gamma}, L)} 
\int d \beta \sin\beta \, d^L_{22}(\beta) 
\biggl\{ \langle 0|b^{N-1} b_{2} \hat O e^{-i \beta L_y} 
(b^{\dagger})^{N-1} b_{2}^{\dagger} | 0\rangle  \nonumber\\
&&\hskip 6cm + 2 \xi_{\gamma} \langle 0|b^{N-2} b_1^2 \hat O e^{-i \beta L_y} 
(b^{\dagger})^{N-1} b_2^\dagger | 0\rangle \nonumber\\
&&\hskip 6cm + \xi_{\gamma}^2
\langle 0|b^{N-2} b_{1}^2 \hat O e^{-i \beta L_y} (b^{\dagger})^{N-2}
(b_{1}^{\dagger})^2 | 0\rangle \biggl\},
\label{gamo}
\end{eqnarray}
where ${\cal N}(\phi_\gamma, L)$ is the normalization for the $\gamma$ band, 
obtained from eq.~(\ref{gamo}) using the identity operator for $\hat O$.
The contribution from the orthogonality terms to the band energies are of the
order $1/N^2$, and therefore they were ignored in the original papers
\cite{kuy88}. In description of high-spin states, however, these terms make
essential contributions and they have to be included in the calculations.
Each contraction of the intrinsic boson operators in (\ref{gamo}) leads to
projected single-boson overlaps of the form $x_{lm} x_{lm'} d^l_{mm'}$. The
resulting Wigner $d$-functions are coupled to a final $d$-function to perform
the $\beta$ integral. This process leads to rather long expressions for the
orthogonality terms. In order to reduce their size, we introduce a compact
notation for the recoupling coefficients as follows 
\begin{eqnarray}
&&\hskip -1cm R_2(jmm',lnn';I) = x_{jm} x_{jm'} x_{ln} x_{ln'} 
\langle jm ln|I m+n \rangle \langle jm' ln'|I m'+n' \rangle, \nonumber\\
&&\hskip -1cm R_3(jmm',lnn',k\mu \mu';I,J) = R_2(jmm',lnn';I) 
x_{k\mu} x_{k\mu'} \langle I m+n k\mu|J m+n+\mu \rangle \nonumber\\
&&\hskip 4cm \langle I m'+n' k\mu'|J m'+n'+\mu' \rangle, \nonumber\\
&&\hskip -1cm R_4(jmm',lnn',k\mu \mu',k'\nu \nu';I,I',J) = R_2(jmm',lnn';I)
R_2(k\mu \mu',k'\nu \nu';I') \nonumber\\
&&\hskip 5.5cm \langle I m+n I'\mu+\nu|J m+n+\mu+\nu \rangle \nonumber\\
&&\hskip 5.5cm \langle I m'+n' I'\mu'+\nu'|J m'+n'+\mu'+\nu' \rangle.
\label{r}
\end{eqnarray}
Higher recoupling coefficients ($R_5, R_6$) are defined similarly.
Using this notation, the reduced normalization for the $\gamma$ band,
$F_\gamma(N,L) = {\cal N} (\phi_\gamma, L)/(2L+1)$, can be written as
\begin{eqnarray}
&&\hskip -1cm F_\gamma(N,L) = \sum_{Ij}\langle L2 j-2|I0\rangle^2  
\nonumber\\
&&\hskip 1cm \times\biggl\{ x_{j2}^2 F(N-1,I)  
+ (N-1) \sum_{ll'} R_2(l'20,l02;j) F(N-2,I) \nonumber\\
&&\hskip 1cm + 2 \xi_\gamma \biggl[ 2\sum_{ll'}  R_2(l'21,l01;j) F(N-2,I) 
\nonumber\\
&&\hskip 2cm + (N-2) \sum_{kk'll'} R_3(k10,k'10,l02;l',j)  F(N-3,I) \biggr] 
\nonumber\\
&&\hskip 1cm + {2\xi_\gamma^2 \over N-1} \biggl[ \sum_{ll'} R_2(l'11,l11;j) 
F(N-2,I) \nonumber\\
&&\hskip 2.5cm + 2(N-2) \sum_{kk'll'} R_3(k10,k'01,l11;l',j) F(N-3,I) 
\nonumber\\
&&\hskip 2.5cm + {1\over 2}(N-2)(N-3) \sum_{kk'k''ll'l''} 
R_4(k10,k'10,l01,l'01;k'',l'',j) \nonumber\\
&&\hskip 7cm \times F(N-4,I) \biggr] \biggr\}.
\label{gamnorm}
\end{eqnarray}
Eq.~(\ref{gamnorm}) expresses the $\gamma$-band normalization in terms of the
ground-band normalization (\ref{norm2}), and it can be evaluated to
any order in $1/N$ using Mathematica. 

The expectation value of the number operator in the 
$\gamma$ band can be calculated similarly, giving
\begin{eqnarray}
&&\hskip -.5cm \langle\hat n_l\rangle_{\gamma,L} =
{1 \over  F_\gamma(N,L)} \sum_{Ij}\langle L2 j-2|I0\rangle^2
\biggl\{ x_{l2}^2 \delta_{jl} F(N-1,I)  \nonumber\\
&&\hskip 1cm + (N-1) \sum_{l'} \biggl(R_2(l'22,l00;j) + 2R_2(l'20,l02;j)
\biggr) F(N-2,I) \nonumber\\
&&\hskip 1cm + (N-1)(N-2) \sum_{kk'l'} R_3(k20,k'02,l00;l',j) F(N-3,I) 
\nonumber\\
&&\hskip 1cm + 2 \xi_\gamma \biggl[ 2\sum_{l'}  \biggl( R_2(l'21,l01;j) 
+ R_2(l'01,l21;j)\biggr) F(N-2,I) \nonumber\\
&&\hskip 2cm + (N-2) \sum_{kk'l'} \biggl( 2R_3(k21,k'01,l00;l',j) 
+ 2R_3(k20,k'01,l01;l',j) \nonumber\\ 
&&\hskip 4.5cm + R_3(k10,k'10,l02;l',j) \biggr) F(N-3,I) 
\nonumber\\
&&\hskip 2cm + (N-2)(N-3) \sum_{kk'k''l'l''} R_4(k10,k'10,l'02,l00;k'',l'',j) 
F(N-4,I) \biggr] \nonumber\\
&&\hskip 1cm + {2\xi_\gamma^2 \over N-1} \biggl[ 2 \sum_{l'} 
R_2(l'11,l11;j) F(N-2,I) \nonumber\\
&&\hskip 2.5cm + (N-2) \sum_{kk'l'} \biggl(R_3(k11,k'11,l00;l',j) 
+ 4R_3(k11,k'10,l01;l',j) \nonumber\\
&&\hskip 5cm + 2R_3(k10,k'01,l11;l',j) \biggr) F(N-3,I) \nonumber\\
&&\hskip 2.5cm + 2(N-2)(N-3) \sum_{kk'k''l'l''} 
\biggl(R_4(k11,k'10,l'01,l00;k'',l'',j) \nonumber\\
&&\hskip 4cm + R_4(k10,k'10,l'01,l01;k'',l'',j) \biggr) F(N-4,I) \biggr] 
\biggr\}.
\label{gam1b}
\end{eqnarray}
It can be easily checked that the condition, 
$\sum_l \langle\hat n_l\rangle_{\gamma,L}=N$ is satisfied by (\ref{gam1b}).
The expectation value of a general two-body interaction in the $\gamma$ band is
given by
\begin{eqnarray}
&&\hskip -1cm \langle T^k \cdot T^k\rangle_{\gamma,L} =  
\sum_{jl} {2k+1 \over 2l+1} t_{kjl}^2 \langle\hat n_l\rangle_{\gamma,L}
\nonumber\\
&&\hskip 1.5cm + {(N-2) (2k+1) \over F_\gamma(N,L)} \sum_{jlj'l'J} 
t_{kjl} t_{kj'l'} 
\left\{ \begin{array}{ccc} j'&j&J \\ l&l'& k\end{array}\right\}
\sum_{iII'} \langle L2I'-2|I0 \rangle^2 \nonumber\\
&& \times \biggl\{ (N-1) \biggl[ {2\over N-2} 
\bigr( P_{2002} + P_{0022}  \bigr) F(N-2,I) \delta_{i0} \nonumber\\
&&\hskip 2cm + \biggr( x_{i2}^2 P_{0000} + 4 x_{i2} x_i P_{2000} \biggr) 
F(N-3,I) \nonumber\\
&&\hskip 2cm + (N-3) \sum_{kk'} R_2(k20,k'02;i) P_{0000} F(N-4,I) \biggr] 
\nonumber\\
&& + 4\xi_\gamma \biggl[ {2\over N-2} P_{2101} \delta_{i0} F(N-2,I) 
\nonumber\\
&&\hskip 1cm + \biggl( 2x_{i2} x_{i1} P_{0001} +  x_{i2} x_i P_{0101} 
+ 2x_{i1} x_i (P_{2001} + P_{0021}) \biggr) F(N-3,I) \nonumber\\
&&\hskip 1cm + (N-3) \sum_{kk'} \biggl( R_2(k21,k'01;i) P_{0000} + 
2R_2(k20,k'01;i) P_{0001} \nonumber\\
&&\hskip 3.5cm + R_2(k10,k'10;i) P_{0002} \biggr) F(N-4,I)\nonumber\\
&&\hskip 1cm +{1 \over 2} (N-3)(N-4) \sum_{kk'k''i'} R_3(k10,k'10,i'02;k'',i) 
P_{0000} F(N-5,I) \biggr] \nonumber\\
&& +{4\xi_\gamma^2 \over N-1} \biggl[ {1\over N-2} P_{1111} 
F(N-2,I) \nonumber\\
&&\hskip 1.5cm + 2\biggl (x_{i1}^2 (P_{1001} + P_{0011}) + 2x_{i1} x_i 
P_{1011} \biggr) F(N-3,I) \nonumber\\
&&\hskip 1.5cm + (N-3) \sum_{kk'} \biggl( {1\over 2} R_2(k11,k'11;i) P_{0000} 
+ 4R_2(k11,k'10;i) P_{0001} \nonumber\\
&&\hskip 2.5cm + 2R_2(k10,k'01;i)  (P_{1001} + P_{0011})
+ R_2(k10,k'10;i) P_{0101} \biggr) F(N-4,I) \nonumber\\
&&\hskip 1.5cm + (N-3)(N-4) \sum_{kk'k''i'} \biggl( R_3(k11,k'10,i'01;k'',i) 
P_{0000} \nonumber\\
&&\hskip 5.5cm + 2R_3(k10,k'10,i'01;k'',i) P_{0001} \biggr) F(N-5,I) \nonumber\\
&&\hskip 1.5cm + {1\over 4}(N-3)(N-4)(N-5) \sum_{kk'k''i'i''l''} 
R_4(k10,k'10,i'01,i''01;k'',l'',i) \nonumber\\
&&\hskip 7cm \times P_{0000} F(N-6,I) \biggr] \biggr\}.
\label{gam2b}
\end{eqnarray}
Here we have introduced the compact notation for the two-boson m.e.
\begin{eqnarray}
&&P_{m'nmn'} = x_{jm} x_{j'm'} x_{ln} x_{l'n'} 
\langle j'm' jm|J m+m' \rangle \langle ln l'n'|J n+n' \rangle \nonumber\\
&&\hskip 2cm \langle J m+m'\, i 2-m-m'|I'2 \rangle 
\langle J n+n'\, i 2-n-n'|I'2 \rangle.
\label{p}
\end{eqnarray}
The dummy summation indices $j,j',l,l',J,i,I'$ in $P$ are suppressed for
convenience. The first term in eq.~(\ref{gam2b}) is the effective one-body term
that arises from normal ordering of the boson operators in the multipole
interaction, and it is expressed using eq.~(\ref{gam1b}). 

Eqs. (\ref{gam1b},\ref{gam2b}) are the counterparts of eqs. 
(\ref{nl2},\ref{tk1}) for the ground band and can be fed directly into
Mathematica for evaluation. As one can surmise from a cursory comparison of 
the parent equations, the resulting third layer expressions are pages long. 
They are not as accurate as the ground band results, presumably
requiring inclusion of even higher order terms. For these reasons, we have
opted for a numerical evaluation of eqs. (\ref{gam1b},\ref{gam2b}) in the
applications. Such a calculation includes all orders in $1/N$, and hence
provides more reliable results for the $\gamma$-band energies.

\subsection{E2 transitions}
Description of the yrast E2 transitions at high-spins is one of the main 
aims of this work. Therefore, we present a brief review of the currently
available $1/N$ results for E2 m.e. and discuss their extensions to higher
orders. A comprehensive study of the E2 transitions among the ground, $\gamma$
and $\beta$ bands was given previously \cite{kuy90}. The first layer m.e.
obtained in \cite{kuy90} for the yrast E2 transitions appears to work rather
well even at high-spins \cite{kuy95a}. Inclusion of the d-boson energy leads to
some deterioration at high-spins, which can be rectified by incorporating
the higher order terms in the expansion. 

The ground band m.e. of the quadrupole operator is given by
\begin{eqnarray}
&&\hskip -1cm \langle L' \parallel Q \parallel L \rangle =
\hat L [4F(N,L') F(N,L)]^{-1/2} \sum_M \langle LM 2-M|L' 0\rangle 
\nonumber\\
&&\hskip 2cm \times \int d\beta \sin \beta\, d^L_{M0} 
\langle 0| b^N Q_{-M} e^{-i \beta L_y} (b^\dagger)^N |0 \rangle,
\label{q1}
\end{eqnarray}
where $\hat L = [2L+1]^{1/2}$. As before, this can be reduced to the form
\begin{eqnarray}
&&\hskip -1cm \langle L' \parallel Q \parallel L \rangle = 
{\sqrt{5} N \hat L \hat L' \over [ F(N,L') F(N,L)]^{1/2}} 
\sum_{jlJ} q_{jl} x_j x_l \nonumber\\
&&\hskip 2cm \times \langle j0L'0|J0 \rangle \langle L0l0|J0 \rangle
\left\{ \begin{array}{ccc} j&L'&J \\ L&l& 2\end{array}\right\} F(N-1,J),
\label{q2}
\end{eqnarray}
which can be evaluated to any order using Mathematica. Because of the tensor
nature of the E2 operator, however, the resulting third layer expressions are
much more complicated than those for the Hamiltonian. In contrast, 
because there is no variation involved, numerical evaluation of eq.~(\ref{q2}) 
is straightforward and is preferred over the lengthy algebraic forms in the
following.

\subsection{Comparison with the exact results}
Before applying the $1/N$ expansion results, we compare them with those
obtained from an exact diagonalization of the Hamiltonian \cite{li93}.
Of necessity, the boson number is fixed at $N=10$.
The Hamiltonian parameters are as in sect. 3 
($\kappa_2=-20$ keV, $\kappa_4=0$, $q=0.5$, $\eta_d=1.5$, $\eta_g=4.5$),
except where noted. Fig.~3 shows a comparison of the ground band energies
normalized with $\bar L=L(L+1)$ so that they all have the same energy scale. 
Fig.~3a studies the convergence of the $1/N$ results obtained with the
VAP procedure. The second layer $1/N$ results (dotted line) rapidly diverge
from the exact energies (circles) for spins $L>2N$, and hence are not
reliable in applications to high-spin states. The third layer results
(dashed line), on the other hand, track the exact energies within a few percent
up to the maximum spin $L=4N$. Using the numerical technique described in 
Appendix A, one can evaluate the $1/N$ expansion to all orders (solid line)
which exhibits an almost perfect agreement with the exact energies. 
This study demonstrates that the third layer $1/N$ expansion results are both
necessary and sufficient for a reliable description of high-spin sates.
Fig.~3b shows the effect of the one-body energies on the accuracy of the third
layer results. The top line is as in fig.~3a, the middle one compares
the exact and $1/N$ results for $\eta_d=0$, $\eta_g=4.5$, and the bottom one
for $\eta_d=0$, $\eta_g=0$. It is seen that the agreement for a pure 
quadrupole Hamiltonian is excellent at all spins, while the addition of g-boson
energy leads to a few percent deviation at very high-spins. In a typical
situation with d-boson energy, this few percent deviation starts occurring at
medium high-spins. 

In fig.~4a, we present a similar study for the gamma band energies.
The average behaviour is well reproduced by the $1/N$ expansion
results but staggering is underestimated. This happens because staggering 
is caused mainly by band mixing between the ground and $\gamma$ bands which is
not included in the present calculations (note that the odd-spin levels, which
are not affected by band mixing, are very well reproduced). We have not
attempted to include band mixing effects here because they are strongly
suppressed for the larger $N$ values used in deformed nuclei \cite{kuy90}, and
hence they can be ignored for the purposes of this work. 
Finally, in fig.~4b, we compare the $1/N$ results for the yrast $E2$ transition
m.e. with the exact ones (circles). The dashed line shows the first
layer result obtained from eq.~(\ref{e2}) which is accurate to a few percent
for $L<2N$, but progressively gets worse with increasing spin. 
Solid line shows the numerical evaluation of eq.~(\ref{q2}), which is complete
to all orders in $1/N$. The agreement with the exact results becomes almost
perfect in this case, including the highest spins which are dominated by $g$
bosons.

\section{Systematic studies \label{sys}}
Application of the IBM to deformed nuclei has been criticized on a number
of fronts ranging from the energy scale of $\gamma$ excitation modes
to the spin dependence of MOI and E2 transitions \cite{bm82}. 
Most of these problems have been attributed to the truncation of the model
space to s and d bosons but due to lack of adequate tools they have not been 
properly addressed in the framework of the sdg-IBM. 
The analytic formulas obtained in the last section have the advantage that one
can easily perform systematic studies of key physical quantities and obtain
useful insights on the effect of various parameters. In this section, we
present such a study that will shed light on the above problems and suggest
more appropriate Hamiltonians for description of deformed nuclei. 
As mentioned in sect. \ref{hamiltonian}, the minimal sdg-IBM Hamiltonian with
quadrupole interaction and g-boson energy has previously been applied to
deformed nuclei and found to be inadequate on most counts \cite{kuy91,kuy95b}.
Inclusion of the d-boson energy and hexadecapole interaction appear to be
necessary to improve this situation. Here we focus on the effect of
these extra terms on the minimal Hamiltonian, especially with regard to
high-spin sates.

To simplify the discussion, we rewrite the ground and $\gamma$ band energies as
\begin{equation}
E_{gL} = \lambda_{g1} \bar L + \lambda_{g2} \bar L^2 + \lambda_{g3} \bar L^3, 
\quad 
E_{\gamma L} = E_\gamma + \lambda_{\gamma 1} \bar L 
+ \lambda_{\gamma 2} \bar L^2 + \lambda_{\gamma 3} \bar L^3,
\label{energy}
\end{equation}
where the coefficients $\lambda_n$ can be read from the respective energy 
expressions. Eq.~(\ref{energy}) is the familiar rotational expansion of the
level energies used in the geometrical model \cite{bm75}. 
The difference between the two models is that in the IBM the coefficients
$\lambda_n$ follow from an underlying Hamiltonian (which is used in describing
other properties) whereas in the geometrical model they are directly extracted
from the data. The MOI problem raised in Ref. \cite{bm82} refers to the fact
that i) the $\lambda_1$ coefficient gets a substantial contribution
from the dipole interaction, $L\cdot L$, which has no dynamical content,
ii) the $\lambda_2$ coefficient is much smaller than the experimental values, 
and iii) the variation in $\lambda_1$ among different bands can not be 
described. All three problems are in fact interrelated. Although the second can
be resolved by renormalizing the moment of inertia at high-spins (e.g. by
modifying $L\cdot L \to L\cdot L/(1+f L\cdot L$) \cite{yos91}), such
modifications are purely kinematical in origin and do not address the dynamical
problem. Further quantities of interest in the study of high-spin states are
the yrast E2 transitions. For systematics, it is sufficient to consider the
first layer $1/N$ expansion result which has the generic form 
\begin{equation}
\langle L-2 \parallel T(E2) \parallel L \rangle =
e_2 N \hat L \langle L0\, 20|L-2\, 0\rangle \bigl[m_1 + m_2 L(L-1)\bigr]
\label{e2}
\end{equation}
where the coefficients $m_n$ are given in ref. \cite{kuy90}. The first term in
(\ref{e2}) gives the familiar rigid-rotor result. The second term is negative
and is responsible for the falloffs predicted in E2 transitions.

In presenting systematics, we find it convenient to use ratios
which eliminate the undesired effects of the scale parameters $\kappa_2$ and
$N$. The energy scale can be fixed, for example, by fitting $\kappa_2$ to the 
excitation energy of the $\gamma$ band, $E_\gamma$.
We discuss five such ratios as a function of $q$ for various values of
(a) $\eta_d=\varepsilon_d/N \kappa_2$ and (b) $\zeta_4=\kappa_4/\kappa_2$. 
The parameter q is varied from 0-1 which covers the whole range
of the quadrupole operator from the $\gamma$-unstable to the SU(3) limit.
$\eta_d$ is varied from 0-2 in 10 equal steps, and $\zeta_4$ from 0-0.5 in 5
equal steps, which cover the range of values used in the applications. 
(Negative values of $\zeta_4$ on the whole are found to have an adverse effect
and hence are not considered.) 
In the $\eta_d$ systematics study, $\zeta_4=0$ is used as its precise value
does not have much influence on the results. In the $\zeta_4$ study, however,
the choice of $\eta_d$ does have an impact, and we adapt $\eta_d=1.5$ which is
the average value used in the applications.
The g-boson energy has been found to have negligible effect \cite{kuy95b}, and
therefore is not varied here but fixed at $\eta_g=4.5$. 
Below, we comment on the behaviour of each ratio and contrast them with the
experimental data. For reference, we note that $q$ assumes values around $\sim
0.5$ in the rare-earth nuclei and $\sim 0.7$ in the actinides. 

1) $E_\gamma/N \lambda_{g1}$ (fig.~5): This ratio relates the energy scales of
the $\gamma$ and ground bands, and its mismatch with experiment has been a
source of criticism \cite{bm82}. It is around 4-5 in the rare-earth region and
increases to 8-9 in the actinides. From fig.~5a it is seen that the CQF with
$\eta_d=0$ overestimates it by about a factor of 2 in both regions. It
decreases rapidly with $\eta_d$ though, and through a judicious use of the
d-boson energy, it should be possible to describe this ratio (and hence the
MOI) without appealing to the $L\cdot L$ term. Fig.~5b shows a
similar study on the effect of the hexadecapole interaction which is seen to be
going in the right direction but is too small to have any impact.

2) $N^2 \lambda_{g2}/\lambda_{g1}$ (fig.~6): This ratio measures the deviation
from the rigid rotor behaviour due to loss of pairing. It ranges from about
-0.2 in the rare-earth region to -0.1 in the actinides. The CQF with $\eta_d=0$
gives values an order of magnitude smaller (fig.~6a) and hence signally fails
in accounting for the spin dependence of MOI as first pointed out in ref.
\cite{bm82}. However, this ratio is very sensitive to the $\eta_d$ values and
the experimental range can be easily attained by including the d-boson energy
in the Hamiltonian. From fig.~6b, the hexadecapole interaction is seen to
have a coherent effect in further reducing this ratio away from the rigid rotor
behaviour.

3) $N^2 \lambda_{\gamma 2}/\lambda_{\gamma 1}$ (fig.~7):  An identical study 
for the $\gamma$ band indicates broadly similar but somewhat larger effects 
of the d-boson energy on the behaviour of $\gamma$ band MOI (fig.~7a). 
A softer MOI in the $\gamma$ band is in line with data in most deformed nuclei
though there are a few exceptions as will be seen in the applications. 
The hexadecapole interaction has an opposite effect (fig.~7b) which
reduces the difference between the ground and $\gamma$ band MOI caused by the
d-boson energy. 

4) $\lambda_{\gamma 1}/\lambda_{g1}$ (fig.~8): This ratio compares the MOI of 
ground and $\gamma$ bands which fluctuates within a band of $\pm 10\%$
across the deformed nuclei. The earlier IBM calculations gave results near one
and could not accommodate such fluctuations. Fig.~8a-b shows that inclusion of
the d-boson energy can increase this ratio by up to 20-30\%, while the
hexadecapole interaction can reduce it significantly (up to 20-30\%) thereby
covering the whole range of fluctuations. 

5) $N^2 m_2/m_1$ (fig.~9): As there is no boson cutoff effect, experimentally
this ratio is consistent with zero. The CQF with $\eta_d=0$ leads to values 
similar to the sd-IBM (fig.~9a), hence despite the addition of g bosons, it
would suffer from the same boson cutoff problem. Introduction of the d-boson
energy, however, reduces it substantially, becoming more in line with
experiments. The effect of the hexadecapole interaction on this ratio (fig.~9b)
is similar to fig.~5b; it is positive but comparatively too small to make
a difference.

Another ratio, namely,  $N^4 \lambda_{g3}/\lambda_{g1}$ is also of interest
especially at very high-spins ($L=20-30$) where the cubic term in eq.
(\ref{energy}) plays an important role \cite{kuy95b}. It exhibits a similar
dependence on $\eta_d$ and $\zeta_4$ as  $N^2 \lambda_{g2}/\lambda_{g1}$ 
in fig.~6, so it is not discussed further here.

The d-boson energy has been mostly neglected in studies of deformed nuclei, 
presumably due to the success of the CQF with $\varepsilon_d=0$ in explaining 
the energy and E2 transition systematics of low-lying states \cite{cas88}. 
In fact, for small values ($\eta_d\sim 1)$, its effect on low-lying states is
negligible and it is not really needed in their description \cite{lip85}. 
The CQF, however, basically leads to a rigid MOI and can not explain either
its spin dependence or its variation among different bands. The obvious way
towards a softer energy surface is to include the d-boson energy in the
Hamiltonian which is seen to vastly improve the description of the
spin-dependent terms in the level energies and E2 transitions. 
The hexadecapole interaction performs a similar function but has a much
smaller effect. The exception is, of course, variations in MOI which could not
be reproduced without the hexadecapole interaction.

\section{Applications to deformed nuclei \label{application}}
In the light of the systematic trends discussed above, we carry out fits
to the rare-earth nuclei $^{158-162}$Dy, $^{164-168}$Er, $^{168-176}$Yb,
$^{170-178}$Hf and the actinides $^{228-232}$Th, $^{234-238}$U.
The isotopes chosen are all well deformed rotors with energy ratio $E_4/E_2$
close to 3.3. We have excluded those exhibiting backbending as their proper
description requires inclusion of two-quasiparticle states in the model space.
While we mainly focus on the description of high-spin states, which has
not been done before, we also consider a selected set of low-lying bands.
This is important in properly constraining the model parameters so that
the results obtained are valid in a broader sense and not just for a small
subset of observables. The sdg-IBM parameters used in the fits are listed in
table 1. Each parameter is particularly sensitive to a certain set of
observables which simplifies the fitting process. For example, $q$ is
determined from interband E2 transitions, $\eta_d$ from the spin dependence of
MOI and $E_\gamma$ (cf. figs. 5-6), $\eta_g$ from $E_{3^+}$, $E_{4^+}$,
$\zeta_4$ from MOI variation (cf. fig.~8), and finally $\kappa_2$ from
the overall energy scale of the spectrum. The parameters are either
constant in a given isotope chain or change smoothly in accordance with
the vibration-rotation shape transition, e.g., $\eta_d$
decreases with increasing $N$ as the nuclei considered become more
rotational. 

The representative observables chosen to describe the low-lying band structure
are the band excitation energies $E_\beta$, $E_\gamma$, $E_{3^+}$, $E_{4^+}$
(table 2), the interband E2 m.e. for $2_{\beta,\gamma} \to 0_g$ transitions
(table 3), and the E4 m.e. for $4_{\gamma,3^+,4^+} \to 0_g$ transitions
(table 4). Note that the E4 m.e. are normalized with the ground transition, so
that an effective E4 charge is not needed in table 1.
With a few exceptions to be discussed below, the general trends of the 
$\beta$ and $\gamma$ band systematics are well reproduced by the 
calculations. The sudden fluctuations seen in some of the band-head energies
(table 2) can be accommodated by a careful tuning of the parameters. 
Since our aim here is to delineate the systematic features of high-spin
states, rather than to obtain refined fits to individual nuclei, we have not
attempted such an improvement. Description of interband E2 transitions is one
of the strong points of the IBM, and as can be seen from table 3, they are
very well reproduced using almost constant $q$ values. 
The parameters in the E4 operator are determined from the conditions 
(\ref{hexpar}), hence the E4 m.e. ratios presented in table 4 are parameter
free predictions of the model. Again the overall agreement with the data is
reasonable which gives confidence on the choice of the E4 operator.
The quality of agreement obtained in tables 2-4 indicates that the 
limited set of sdg-IBM parameters (table 1) can describe the basic features 
of the low-lying bands in deformed nuclei. 

In the study of high-spin sates, we include the level energies for the ground
and $\gamma$ bands, and the yrast E2 transitions for each set of isotopes
(figs. 10-27). We first comment on their general features. In all cases, the
MOI strongly deviates from the rigid rotor behaviour which would 
be represented by a horizontal line in the figures. Further, this deviation is 
not linear but curves up with increasing spin underscoring the importance of
the cubic term in eq.~(\ref{energy}). Note that because of the ample data 
available, these features are most clear in the ground bands and to a lesser 
extent in the $\gamma$ bands. The yrast E2 m.e., on the other hand,
follow closely the rigid rotor values with no sign of a boson cutoff effect. As
emphasized in sect. 5, these properties can be explained in the IBM by
including the d-boson energy in the Hamiltonian. Below we comment on the
specific features of each isotope chain. 

1) $^{158-162}$Dy (figs. 10-12):  
Among the deformed nuclei considered in this work, $^{158}$Dy, together with
$^{170}$Hf, exhibit the largest changes in MOI. These nuclei have the lowest
boson numbers among the rare-earth set and are clearly influenced by 
the vibration-rotation phase transition as indicated by the larger $\eta_d$
values used. The ground band energies (fig.~10) are well described with
relative errors of about 1-2\%. The trend in $\gamma$ band energies (fig.~11)
is similarly reproduced (note the different scales in figs. 10 and 11). 
The slight overprediction of energies here can be improved by fine tuning the
hexadecapole interaction (cf. fig.~8b). The yrast E2 m.e. have been a sore
point in applications of the sd-IBM to high-spin states due to boson cutoff.
For example, in $^{158}$Dy, the sd-IBM would predict band termination at $L=26$
which is not seen in the data (fig.~12a). This problem has been resolved in the
present sdg-IBM calculations which account for the yrast E2 data very well
(fig.~12). A side remark for $^{162}$Dy is that the band excitation energies
in this nucleus do not follow the trend of $^{158-160}$Dy (table 2), hence
it requires individual attention for a better description.

2) $^{164-168}$Er (figs. 13-15): 
The Er isotopes, and in particular $^{168}$Er, are the exceptional cases
mentioned above for which a consistent description of the data could not be
obtained with our limited set of parameters. While the spin dependence of the
ground and  $\gamma$ band MOI (figs. 13-14) and the E2 m.e. (fig.~15, table 3)
are well described, the band excitation energies are overpredicted (table 2)
and the E4 m.e. are rather poor (table 4). The problem stems from the fact
that, among all the deformed nuclei considered in this study, the Er isotopes
have the lowest lying $\gamma$ bands and the most rigid MOI. As seen from figs.
6-7, these two quantities are correlated in the present parametrization, so
that a lower $\gamma$ band obliges a softer MOI (cf. figs. 6-7). Thus a proper
description of the Er isotopes requires extension of the Hamiltonian
(\ref{hamsdg}) and/or relaxation of the constraints on the quadrupole and
hexadecapole parameters. For example, in a detailed study of $^{168}$Er in the
sdg-IBM, 14 parameters were employed \cite{yos88}. Here we will be content with
exposing the exceptional nature of the Er isotopes and leave their detailed
investigation for future work. 

3) $^{168-176}$Yb (figs. 16-18): 
The Yb isotopes are uniformly well described and require little comment. In
contrast to Er, the $\gamma$ band energies in the Yb isotopes are higher which
are well correlated with their relatively stiff MOI (fig.~16). One point
worthwhile to make is that the $\gamma$ band MOI is higher than that of the
ground band which could not be explained without the hexadecapole interaction. 

4) $^{170-178}$Hf (figs. 19-21): 
In the Hf isotopes, the $\gamma$ band comes down but the MOI are softer, and
hence the correlation between the two quantities is preserved (fig.~19). The
staggering observed in the $\gamma$ bands (fig.~20) requires inclusion of band
mixing effects for a better description. Otherwise the data are well reproduced
by the calculations. 

5) $^{228-232}$Th (figs. 22-24): 
Although boson numbers are lower in the actinide nuclei considered here, they
exhibit characteristics of well deformed nuclei. The MOI in actinides are
typically twice as large as those in rare-earths requiring smaller $\kappa_2$
values. The high-spin data are scarce in $^{228-230}$Th but in $^{232}$Th,
where data up to spin $L=30$ are available, an excellent description is
obtained. One interesting feature of the $\gamma$ band MOI in $^{232}$Th is
that it is larger and stiffer compared to the ground band. Both of these
features require a large dose of hexadecapole interaction for explanation as
remarked in the systematic studies. 

6) $^{234-238}$U (figs. 25-27): 
The most extensive high-spin data are available for the yrast bands in the
U isotopes which are well described by the present calculations. The yrast E2
m.e. in the U isotopes (and $^{232}$Th) were measured to check the boson cutoff
predictions of the sd-IBM, i.e. E2 m.e. vanish at $L=2N$. As seen in fig.~27,
the E2 data show no sign of falloff which provides one of the strongest
motivations for inclusion of g bosons. At the highest spins, the sdg-IBM
calculations appear to underpredict the E2 measurements. We emphasize that this
is not due to any boson cutoff effect but rather due to deviation of the data
from the rigid rotor values. That is, most models would have difficulty in
explaining these E2 transition m.e. which are larger than the rigid rotor
values (see, for example, \cite{tro94}).

\section{Summary and conclusions}
In this paper, we have presented a systematic description of high-spin states
in deformed nuclei within the framework of the sdg-IBM. Such a description has
been long overdue but has not been performed earlier due to the technical
difficulties in diagonalizing the sdg-IBM Hamiltonians. We have shown that the
$1/N$ expansion formalism, extended to higher orders using computer algebra,
provides a viable alternative for this purpose. Systematic studies of the model
parameters using the $1/N$ expansion formulas have indicated that some of the
long standing problems associated with the description of MOI and E2
transitions in the IBM can be resolved by including the d-boson energy in the
Hamiltonian. The hexadecapole interaction has a minor effect on the ground band
but could play a decisive role on $\gamma$ and other excited band properties. 
This feature of the hexadecapole interaction does not appear to be well 
appreciated in literature. For example, the recently observed staggering effect
in some superdeformed bands  have been attributed to the hexadecapole degrees
of freedom. If this is true, then it would have profound effects on the 
neighbouring non-yrast bands. 

The application of the sdg-IBM Hamiltonian consisting of single-boson energies,
and quadrupole and hexadecapole interactions (with constrained parameters)
resulted in a mostly uniform and successful description of both low-lying
band structures and high-spin states across the rare-earth and actinide
regions. In the past, many experimental results on high-spin states were
compared with the sd-IBM calculations with negative connotations. This was
presumably due to the lack of the sdg-IBM calculations. We hope that the
extensive sdg-IBM results presented in this work will help remedying this
situation. 

Our limited parametrization did not work as well in the case of the Er isotopes
which appear to have rather exceptional properties as exposed in this 
systematic study. We are planning to carry out a more detailed study of
the Er isotopes without imposing the constraints used in this paper.
It is amusing to note that $^{168}$Er, which has been used as a benchmark case
in tests of phenomenological models, may turn out to be the most exceptional of
all deformed nuclei.

\section*{Appendix A}
In cases where the higher order $1/N$ expressions are too complicated to be
useful, numerical evaluation of the $1/N$ expansion formulas may provide
a more practical alternative.
The reduced normalization integral (\ref{norm1}) can be written explicitly as
\begin{equation}
F(N,L) =  \int_0^1 dz  P_L(z) 
\bigg[ \sum_{l} x_l^2 P_l(z)\biggr]^N,
\label{a1}
\end{equation}
where $P_L(z)$ is a Legendre polynomial.
In the sd-IBM, the term in brackets is given by $(c_0 + c_2 z^2)^N$, and
in the sdg-IBM by $(c_0 + c_2 z^2 + c_4 z^4)^N$ where $c_i$ are constants
involving the mean fields.
By expanding the binomial or trinomial, and using
\begin{equation}
\int_0^1 dz P_L(z) z^n = {n! \over (n-L)!! (n+L+1)!!},
\end{equation}
eq.~(\ref{a1}) can be integrated term by term.
This has the advantage that $F(N,L)$ and hence $\langle H \rangle$ are
calculated exactly to all orders in $1/N$. The variational problem is solved
numerically using the exact m.e. (\ref{nl2},\ref{tk1}) and the simplex method.
The numerical evaluation can be used, for example, to check the convergence
of $1/N$ expressions and determine whether a calculation to a particular 
order is accurate enough.

\eject

\vfill \eject
\noindent
{\Large \bf Figure captions}
\\[.4cm]
Fig. 1. Effect of the basis space truncation on a) band excitation energies
b) E2 transitions, and c) E4 transitions. The maximum number of g bosons,
$n_{g\max}$, allowed in the basis space is increased from 1 to the maximum of
$N=10$. Parameters of the sdg-IBM Hamiltonian are given in the text.
\\[.4cm]
Fig. 2. Effect of the basis space truncation on the ground band a) excitation 
energies, and b) E2 transitions. 
\\[.4cm]
Fig. 3. Comparison of the ground band energies obtained from the 1/$N$
expansion with the exact diagonalization results (circles). In (a) different
lines refer to the second layer calculation (dotted line), the third layer
(dashed line), and the numerical one to all orders (solid line). In (b) the
lines correspond to the third layer results obtained with $\eta_d=1.5$,
$\eta_g=4.5$ (top), $\eta_d=0$, $\eta_g=4.5$ (middle), $\eta_d=0$, $\eta_g=0$
(bottom). 
\\[.4cm]
Fig. 4. (a) Comparison of the $\gamma$ band energies obtained from the 1/$N$
expansion with the exact diagonalization results (circles).
(b) Comparison of the yrast E2 transition m.e. obtained from eq.~(\ref{e2}) 
(dashed line) with the exact diagonalization results (circles). The solid line
shows the numerical evaluation of the m.e. to all orders.
\\[.4cm]
Fig. 5. The effect of (a) the d-boson energy and (b) the hexadecapole
interaction on the ratio  $E_\gamma/N \lambda_{g1}$ which relates the energy
scales of the $\gamma$ and ground bands.
\\[.4cm] 
Fig. 6. Same as fig. 5 but for the ratio $N^2 \lambda_{g2}/\lambda_{g1}$
which measures the deviation of the ground band moment of inertia from the
rigid rotor behaviour. 
\\[.4cm] 
Fig. 7. Same as fig. 6 but for the $\gamma$ band.
\\[.4cm]
Fig. 8. Same as fig. 5 but for the ratio $\lambda_{\gamma 1}/\lambda_{g1}$
which compares the moment of inertia of the ground and $\gamma$ bands.
\\[.4cm]
Fig. 9. Same as fig. 5 but for the ratio $m_2/m_1$ which measures the boson
cutoff effect.
\\[.4cm]
Fig. 10. Comparison of the experimental (circles) and calculated (solid lines)
ground band energies $E_{gL}/L(L+1)$ (in keV) in $^{156-160}$Dy.
The data are from \cite{nds158,nds160,nds162}.
\\[.4cm]
Fig. 11. Same as fig. 10 but for the $\gamma$ bands.
\\[.4cm]
Fig. 12. Comparison of the experimental (circles) and calculated (solid lines)
yrast $E2$ transitions in $^{156-160}$Dy. The data are from 
\cite{nds158,nds160,nds162}.
\\[.4cm]
Fig. 13. Same as fig. 10 but in $^{164-168}$Er. The data are from 
\cite{nds164,nds166,nds168}
\\[.4cm] 
Fig. 14. Same as fig. 11 but in $^{164-168}$Er. The data are from 
\cite{nds164,nds166,nds168}.
\\[.4cm]
Fig. 15. Same as fig. 12 but in $^{164-168}$Er. The data are from 
\cite{nds164,nds166,nds168}.
\\[.4cm]
Fig. 16. Same as fig. 10 but in $^{168-176}$Yb. The data are from 
\cite{nds168,nds170,nds172,nds174,nds176}.
\\[.4cm]
Fig. 17. Same as fig. 11 but in $^{168-176}$Yb. The data are from 
\cite{nds168,nds170,nds172,nds174,nds176}.
\\[.4cm]
Fig. 18. Same as fig. 12 but in $^{168-176}$Yb. The data are from 
\cite{nds168,nds170,nds172,nds174,nds176}.
\\[.4cm]
Fig. 19. Same as fig. 10 but in $^{170-178}$Hf. The data are from 
\cite{nds170,nds172,nds174,nds176,nds178}.
\\[.4cm]
Fig. 20. Same as fig. 11 but in $^{170-178}$Hf. The data are from 
\cite{nds170,nds172,nds174,nds176,nds178}.
\\[.4cm]
Fig. 21. Same as fig. 12 but in $^{170-178}$Hf. The data are from 
\cite{nds170,nds172,nds174,nds176,nds178}.
\\[.4cm]
Fig. 22. Same as fig. 10 but in $^{228-232}$Th. The data are from 
\cite{nds228,nds230,nds232}.
\\[.4cm]
Fig. 23. Same as fig. 11 but in $^{228-232}$Th. The data are from 
\cite{nds228,nds230,nds232}.
\\[.4cm]
Fig. 24. Same as fig. 12 but in $^{228-232}$Th. The data are from 
\cite{nds228,nds230,nds232}.
\\[.4cm]
Fig. 25. Same as fig. 10 but in $^{234-238}$U. The data are from 
\cite{nds234,nds232,nds238}.
\\[.4cm]
Fig. 26. Same as fig. 11 but in $^{234-238}$U. The data are from 
\cite{nds234,nds232,nds238}.
\\[.4cm]
Fig. 27. Same as fig. 12 but in $^{234-238}$U. The data are from 
\cite{nds234,nds232,nds238}.

\vfill \eject
\begin{table}
\caption{Parameters used in the sdg-IBM calculations. $\kappa_2$ is in
keV and $e_2$ in $e$b.}
\label{table1}
\begin{tabular}{cccccccc} \hline
Nucleus & $N$ &  $\kappa_2$ & $\zeta_4$  & $q$ & $\eta_d$ &
$\eta_g$ & $e_2$ \\ 
\hline
$^{158}$Dy & 13 & 19.8 & 0.30& 0.50 & 1.90 & 5.0 & 0.13 \\
$^{160}$Dy & 14 & 19.9 & 0.30& 0.50 & 1.77 & 4.6 & 0.13 \\
$^{162}$Dy & 15 & 19.3 & 0.35& 0.50 & 1.60 & 4.5 & 0.13 \\
$^{164}$Er & 14 & 22.0 & 0.40& 0.50 & 1.50 & 5.0 & 0.14 \\
$^{166}$Er & 15 & 21.2 & 0.40& 0.50 & 1.42 & 4.7 & 0.13 \\
$^{168}$Er & 16 & 23.1 & 0.40& 0.50 & 1.22 & 4.6 & 0.13 \\
$^{168}$Yb & 14 & 19.7 & 0.35& 0.50 & 1.68 & 5.0 & 0.14 \\
$^{170}$Yb & 15 & 20.6 & 0.35& 0.50 & 1.63 & 4.7 & 0.13 \\
$^{172}$Yb & 16 & 20.5 & 0.25& 0.60 & 1.78 & 5.0 & 0.13 \\
$^{174}$Yb & 17 & 22.2 & 0.25& 0.60 & 1.66 & 4.1 & 0.12 \\
$^{176}$Yb & 16 & 20.1 & 0.35& 0.60 & 1.83 & 5.3 & 0.12 \\
$^{170}$Hf & 13 & 19.1 & 0.10& 0.50 & 2.04 & 4.9 & 0.14 \\
$^{172}$Hf & 14 & 19.5 & 0.10& 0.50 & 1.99 & 4.5 & 0.13 \\
$^{174}$Hf & 15 & 20.7 & 0.10& 0.50 & 1.91 & 4.1 & 0.13 \\
$^{176}$Hf & 16 & 22.0 & 0.10& 0.50 & 1.80 & 3.5 & 0.13 \\
$^{178}$Hf & 15 & 21.8 & 0.10& 0.50 & 1.83 & 4.1 & 0.13 \\
\hline
$^{228}$Th& 10 & 19.7 & 0.30& 0.68  & 1.60 & 3.0 & 0.20 \\
$^{230}$Th& 11 & 15.5 & 0.40& 0.68  & 1.59 & 4.4 & 0.20 \\
$^{232}$Th& 12 & 14.6 & 0.40& 0.68  & 1.58 & 4.4 & 0.20 \\
$^{234}$U & 13 & 14.9 & 0.20& 0.70 & 1.62 & 3.5 & 0.18 \\
$^{236}$U & 14 & 16.4 & 0.20& 0.70 & 1.57 & 3.2 & 0.17 \\
$^{238}$U & 15 & 17.7 & 0.20& 0.70 & 1.56 & 2.8 & 0.17 \\ 
\hline
\end{tabular}
\end {table}

\begin{table}
\caption{Comparison of the $\beta$, $\gamma$, and $K=3^+, 4^+$ (single-phonon)
band energies (in keV) with the sdg-IBM calculations in the rare-earth 
and actinide regions. The data are from Nucl. Data Sheets.} 
\label{table2}
\begin{tabular}{cccccccccccccc} \hline
 &&\multicolumn{2}{c}{$E_\beta$}
&&\multicolumn{2}{c}{$E_\gamma$}
&&\multicolumn{2}{c}{$E_{3^+}$}
&&\multicolumn{2}{c}{$E_{4^+}$} \\[0.1cm]
\cline{3-4} \cline{6-7} \cline{9-10} \cline{12-13}
Nucleus &&\multicolumn{1}{c}{Cal.}&\multicolumn{1}{c}{Exp.}&&
   \multicolumn{1}{c}{Cal.}&\multicolumn{1}{c}{Exp.}&&
   \multicolumn{1}{c}{Cal.}&\multicolumn{1}{c}{Exp.}&&
   \multicolumn{1}{c}{Cal.}&\multicolumn{1}{c}{Exp.}\\ \hline
$^{158}$Dy&& 916 & 991 && 965 & 946 && 1461  & - && 1935  & 1895 \\
$^{160}$Dy&& 1204 & 1275 && 998 & 966 && 1512  & - && 2085  & - \\
$^{162}$Dy&& 1284 & 1205 && 1048 & 888 &&1617   & - && 2201  & 1536 \\
$^{164}$Er&& 1233 & 1246 && 1145 & 860&& 1634  & 1702 && 2003  & -  \\
$^{166}$Er&& 1275 & 1460 && 1138 & 786 && 1704 & - && 2376  & - \\
$^{168}$Er&& 1590 & 1217 && 1345 & 821 && 1892  & 1654 && 2483 & 2238 \\
$^{168}$Yb&& 1085 & 1156 && 1028 & 984 && 1470 & 1452 && 1706  & - \\
$^{170}$Yb&& 1218 & 1069 && 1138 & 1145 && 1501  & - &&1857   & 1408 \\
$^{172}$Yb&& 1345 & 1043 && 1385 & 1466 && 1799  & 1663 &&2403   & 2073\\
$^{174}$Yb&& 1589 & 1487 && 1554 & 1634 && 1810  & - && 2545  & - \\
$^{176}$Yb&& 1329 & 1779 && 1324 & 1261 && 1685  & - && 2407  & - \\
$^{170}$Hf&& 868 & 880 && 985 & 961 && 1482  & - && 1528  & - \\
$^{172}$Hf&& 982 & 871 && 1016 & 1075 &&1530   & - && 1604  & - \\
$^{174}$Hf&& 1108 & 827 && 1137 & 1227 &&1547   & 1303 && 1669  & - \\
$^{176}$Hf&& 1298 & 1150 && 1278 & 1341 &&1671   & 1578 && 1840  & - \\
$^{178}$Hf&& 1231 & 1199 && 1202 & 1175 &&1710   & 1728 && 1809  & 1848 \\
\hline
$^{228}$Th&& 791 & 832 && 846 & 969 && 1082  & - &&1458   & - \\
$^{230}$Th&& 734 & 635 && 792 & 781 && 1004  & - &&1519   & - \\
$^{232}$Th&& 769 & 730 && 813 & 785 && 1041  & - && 1574  & - \\
$^{234}$U && 825 & 809 && 871 & 927 && 1253  & 1496 && 1566  & 1723 \\
$^{236}$U && 986 & 919 && 1002 & 958 && 1325  & - && 1799  & - \\
$^{238}$U && 1080 & 993 && 1086 & 1060&&1328   & 1059 && 1965  & -  \\
\hline
\end{tabular}
\end {table}

\begin{table}
\label{table3}
\caption{Comparison of the interband E2 transitions (in $e$b) with the 
sdg-IBM calculations in the rare-earth and actinide regions.
The data are from Nucl. Data Sheets.}
\begin{tabular}{cccccccccccccc} \hline
&&\multicolumn{2}{c}{$\langle 2_\beta  ||T(E2)||0_g\rangle$}
&&\multicolumn{2}{c}{$\langle 2_\gamma ||T(E2)||0_g\rangle$} \\[0.1cm]
\cline{3-4} \cline{6-7}
Nucleus &&\multicolumn{1}{c}{Cal.}&\multicolumn{1}{c}{Exp.}&&
   \multicolumn{1}{c}{Cal.}&\multicolumn{1}{c}{Exp.}\\
\hline
$^{158}$Dy&& 0.21 & 0.23 $\pm$ 0.02 && 0.41 & 0.39 $\pm$ 0.04 \\
$^{160}$Dy  && 0.19 & - && 0.34 & 0.27 $\pm$ 0.04 \\
$^{162}$Dy  && 0.17 & - && 0.35 & 0.35 $\pm$ 0.02 \\
$^{164}$Er  &&  0.17 & - && 0.34 & 0.37 $\pm$ 0.02\\
$^{166}$Er  &&  0.16 & - && 0.41 & 0.39 $\pm$ 0.02\\
$^{168}$Er&&  0.15 &$<$0.03 && 0.34 & 0.36 $\pm$ 0.01\\
$^{168}$Yb&&  0.20 & 0.22 $\pm$ 0.01 && 0.35 & 0.36 $\pm$ 0.04 \\
$^{170}$Yb&&  0.19 & 0.17 $\pm$ 0.02 && 0.32 & 0.28 $\pm$ 0.03 \\
$^{172}$Yb&&  0.16 & 0.09 $\pm$ 0.01 && 0.26 & 0.21 $\pm$ 0.03\\
$^{174}$Yb&&  0.14 & - && 0.24 &  0.22 $\pm$ 0.03 \\
$^{176}$Yb&&  0.14 & - && 0.27 &  0.23 $\pm$ 0.03 \\
$^{170}$Hf&&  0.23 & - && 0.36 & - \\
$^{172}$Hf&&  0.19 & - && 0.31 & - \\
$^{174}$Hf&&  0.20 & - && 0.33 & 0.37 $\pm$ 0.04\\
$^{176}$Hf&&  0.18 & 0.21 $\pm$ 0.02 && 0.34 & 0.35 $\pm$ 0.01\\
$^{178}$Hf&&  0.21 & - && 0.32 & 0.34 $\pm$ 0.02\\ 
\hline
$^{228}$Th&& 0.18 & - && 0.28 & - \\
$^{230}$Th&& 0.20 & 0.21 $\pm$ 0.05 && 0.34 & 0.35 $\pm$ 0.06 \\
$^{232}$Th&& 0.23 & 0.31 $\pm$ 0.07 && 0.38 & 0.36 $\pm$ 0.04 \\
$^{234}$U &&  0.21 & $<$0.24 && 0.32 & 0.35 $\pm$ 0.04 \\
$^{236}$U &&  0.20 & - && 0.30 & -  \\
$^{238}$U && 0.19 & 0.23 $\pm$ 0.03&& 0.31 & 0.36 $\pm$ 0.04\\  
\hline
\end{tabular}
\end {table}

\begin{table}
\caption{Comparison of the interband $E4$ transitions, normalized to inband
ones, with the sdg-IBM calculations in the rare-earth and actinide regions.
The data are from Nucl. Data Sheets.}
\label{table4}
\begin{tabular}{cccccccccccccc} \hline
&&\multicolumn{2}{c}{${ <4_{\gamma} ||T(E4)||0_g> \over <4_g||T(E4)||0_g>} $}
&&\multicolumn{2}{c}{${ <4_{3^+} ||T(E4)||0_g> \over <4_g||T(E4)||0_g>}$}
&&\multicolumn{2}{c}{${ <4_{4^+} ||T(E4)||0_g> \over <4_g||T(E4)||0_g>}$} 
\\[0.1cm]
\cline{3-4} \cline{6-7}  \cline{9-10}
Nucleus &&\multicolumn{1}{c}{Cal.}&\multicolumn{1}{c}{Exp.}&&
   \multicolumn{1}{c}{Cal.}&\multicolumn{1}{c}{Exp.}
&&\multicolumn{1}{c}{Cal.}&\multicolumn{1}{c}{Exp.}\\
\hline
$^{158}$Dy&& 0.49 & 0.56 $\pm$ 0.44 && 1.14 & -  &&0.32  &  -\\
$^{160}$Dy&& 0.50 & 0.63 $\pm$ 0.24 && 1.08 & -   &&0.27  & -\\
$^{162}$Dy&& 0.37 & - && 1.06 & -   &&0.28  & -\\
$^{164}$Er&& 0.44 & - && 1.05 & -   &&0.23  & -\\
$^{166}$Er&& 0.53 & - && 1.00 & -   && 0.18 & -\\
$^{168}$Er&& 0.46 & 1.32 $\pm$ 0.72 && 0.95 & 0.60 $\pm$ 0.33  && 0.15 &
0.25 $\pm$ 0.15\\
$^{168}$Yb&& 0.50 & -  && 0.95 & -   && 0.20 & - \\
$^{170}$Yb&& 0.60 & -  && 0.89 & -   && 0.14 &- \\
$^{172}$Yb&& 0.53 & 0.20 $\pm$ 0.19 && 0.82 & 0.69 $\pm$ 0.57   &&0.09 & - \\
$^{172}$Yb&& 0.36 & 0.20 $\pm$ 0.19 && 0.78 & 0.69 $\pm$ 0.57   &&0.16 & - \\
$^{174}$Yb&& 0.44 & - && 0.73 &0.54 $\pm$ 0.29  && 0.11 & - \\
$^{176}$Yb&& 0.37 & - && 0.81 & - && 0.17 & - \\
$^{170}$Hf&& 0.44 & - && 1.17 & -    &&0.43  & -\\
$^{172}$Hf&& 0.45 & - && 1.12 & -   &&0.38  & -\\
$^{174}$Hf&& 0.50 & - && 1.04 & -   && 0.28 & -\\
$^{176}$Hf&& 0.44 & - && 1.01 & -   &&0.28  &- \\
$^{178}$Hf&& 0.43 & - && 1.08 & -   &&0.34  &- \\ 
\hline
$^{228}$Th&& 0.55 & - && 0.62 & - && 0.02 &  -\\
$^{230}$Th&& 0.44 & - && 0.71 & -  && 0.07 & -\\
$^{232}$Th&& 0.48 & - && 0.68 & -  && 0.05 & -\\
$^{234}$U && 0.48 & - && 0.61 & -  && 0.02 & -\\
$^{238}$U && 0.50 & - && 0.58 & - && 0.02 &  -\\
\hline
\end{tabular}
\end {table}

\end{document}